# The identification of the dominant donors in low temperature grown InPBi materials

G. N. Wei[1], D. Xing[2], Q. Feng[1], W. G. Luo[1], Y. Y. Li[2], K. Wang[3], L. Y. Zhang[3], W. W. Pan[3], S. M. Wang[3], S. Y. Yang[1,*] and K. Y. Wang[1,†]

[1] State Key Laboratory of Superlattices and Microstructures, Institute of Semiconductors, Chinese Academy of Sciences, Beijing 100083, China

[2] State Key Laboratory of Semiconductor Materials Science, Institute of Semiconductors, Chinese Academy of Sciences, Beijing 100083, China

[3] State Key Laboratory of Functional Materials for Informatics, Shanghai Institute of Microsystem and Information Technology, Chinese Academy of Sciences, Shanghai 200050, China

**ABSTRACT**

Combined with magnetotransport measurements and first-principles calculations, we systematically investigated the effects of Bi incorporation on the electrical properties of the undoped $InP_{1-x}Bi_x$ epilayers with $0 \leq x \leq 2.41\%$. The Hall-bar measurements reveal a dominant *n*-type conductivity of the InPBi samples. The electron concentrations are found to decrease firstly as $x$ increases up to $x=1.83\%$, and then increase again with further increasing Bi composition, whiles the electron mobility shows an inverse variation to the electron concentration. First-principle calculations suggest that both the phosphorus antisites and vacancy defects are the dominant donors responsible for the high electron concentration. And their defect concentrations show different behaviors as Bi composition $x$ increases, resulting in a nonlinear relationship between electron concentration and Bi composition in InPBi alloys.

Key words: Hall-bar, InPBi alloys, dominant donors, first-principles calculation

## I. INTRODUCTION

III-V semiconductor phosphides are important materials for optoelectronic devices working at visible and near-infrared wavelength ranges [1]. The synthesis of the dilute bismides, by incorporating a small amount of Bi into III-V semiconductors, has attracted great interest in recent years, due to the large band-gap reduction, suppression of the troublesome Auger recombination processes by large spin-orbit splitting, and other unique properties [2-4]. Previous studies have primarily focused on the growth and properties of the GaAs-based bismuthides [5-9], since GaAsBi alloys were firstly synthesized by metalorganic vapor phase epitaxy in 1998 [10] and by molecular beam epitaxy (MBE) in 2003 [11]. Berding et al. theoretically analyzed InSbBi, InAsBi, and InPBi as potential candidates for mid- and far-infrared optoelectronics and pointed out that it is easy to synthesize InSbBi but difficult to grow InPBi [12]. Once formed, InPBi is the most robust one and is expected to be the most attractive candidate among them for narrow-gap applications [12-14]. Experimentally, the InPBi alloy with good single crystal quality has been successfully synthesized recently and exhibits unexpected strong and broad photoluminescence at room temperature [13]. The incorporation of Bi into InP can lead to a large bandgap reduction of 56 meV/Bi%, providing a new way to tune the band structure for potential optoelectronic and electronic applications [13,14]. So far, only a few works have been directed to the understanding of the structural and optical properties of the dilute bismide alloy InPBi with improving quality [13-15]. However, the electrical transport characteristics of the InPBi alloy are poorly understood. In this work, we combine both the electrical transport measurements and first-principles calculations to investigate the Bi content dependent electric properties of the InPBi alloys.

---

* syyang@semi.ac.cn
† kywang@semi.ac.cn

Due to the large size difference and the high tendency of Bi to surface segregate, the InPBi alloys must be grown at low temperatures by gas-source molecular beam epitaxy. Through variable temperature Hall measurements, the nominally undoped InPBi epilayers are found to have very high electron concentration $n_e$ (~$10^{18}$ cm$^{-3}$). Interestingly, the electron concentrations show a nonlinear behavior as the increase of the Bi composition $x$, i.e., $n_e$ decreases with increasing $x$ for $x \leq 1.83\%$ and then increases with increasing Bi doping further, while the electron mobility shows an inverse variation to $n_e$ as a function of Bi compositions. Since the isoelectronic energy level of Bi resides in the valence band of most III-V materials, and it has been theoretically shown that the incorporation of Bi atoms in III-V group materials will introduce impurity levels within valence band [16-18]. Thus the high electron concentration observed in our experiments should not be originated directly from Bi atoms. A similar high electron concentration has been observed in low-temperature (LT) grown InP [19-21], and is believed to originate from the phosphorus antisites $P_{In}$ via auto-ionization with first donor level above the conduction band minimum (CBM) [19,20]. The high electron concentration in InPBi could also originate from native donor defects considering the similar LT grown conditions. However, our first-principles calculations show that the donor level of $P_{In}$ is too deep within the gap, in consistent with previous theoretical studies [22,23]. Besides, the $P_{In}$ leads to a monotonic increasing in electron concentration with increasing $x$ in InPBi. On the other hand, phosphorus vacancy $V_P$ has a donor level above CBM, suggesting $V_P$ should be another dominant donor contributing to the high electron concentration. More importantly, the concentrations of $P_{In}$ and $V_P$ defects are found to show different changes with the increase of Bi composition, which can well explain the nonlinear variation of the electron concentration with increasing the Bi contents in the InPBi alloys.

Our paper is organized as follows. After an introduction in Sec. I, the details of experiments and first-principles calculations are given in Sec. II. The main results from experiments and calculations are presented in Sec. III, and finally a conclusion in Sec. IV.

## II. EXPERIMENT AND CALCULATION DETAILS

The 430 nm thick InP$_{1-x}$Bi$_x$ epilayers were grown on (100) semi-insulating InP substrates by V90 gas source molecular beam epitaxy at growth temperature of around 320 ℃. During the growth process, elemental In and Bi fluxes were controlled by adjusting the respective effusion cell temperatures while $P_2$ was cracked from PH$_3$ at 1000 ℃. The growth rate was 800 nm/h and the PH$_3$ pressure was set to be 350 Torr. Standard photolithography techniques were used to fabricate Hall bars of width 100 μm with neighbor voltage probes separated by 450 μm. Inductively

Coupled Plasma (ICP) dry etching method was used to etch down about 700 nm of the sample, then 20 nm Ti plus 400 nm Au were sputtered as the contacts. Temperature dependence of the conductivity of the devices was carried out at temperatures ranging from 5 to 300 K. The DC measurements were performed for the Hall devices using a fixed electric current of 2 µA, which was small enough to avoid electric heating. The Hall Effect measurements with magnetic field up to 9 T at different temperatures were used to obtain the electron concentrations. All the electric signals were detected by Agilent B1500A, with triaxial cables to avoid parasitic capacitance.

Our first-principles calculations were carried out using the Vienna ab initio simulation package (VASP) [24] with the generalized-gradient approximation of Perdew, Burke, and Ernzerhof (PBE) [25] and the hybrid density functional proposed by Heyd, Scuseria, and Ernzerhof (HSE) [26] for the exchange correlation potential. We adopted the all-electron-like projector-augmented wave potentials [27] and the energy cutoff for the plane-wave expansion was set as 400 eV. A $3\times3\times3$ cubic supercell containing 218 atoms with 1-4 Bi atoms substituting P sites was used to model the InPBi alloys. We assumed the lattice constant of the alloy varies linearly with Bi composition and considered different configurations of complexes of Bi and the defects ($P_{In}$ and $V_P$) in the supercell. Only a single Γ point was used for Brillouin zone integration [28].

The formation energy of defect X in charge state $q$ was calculated by [29]:

$$\Delta H_f(X,q) = E_{tot}(X,q) - E(\text{host}) + n_{In}\mu_{In} + n_P\mu_P + n_{Bi}\mu_{Bi} + q(E_F + E_{VBM}),$$

where $E_{tot}(X,q)$ is the total energy of the supercell containing the defect X, and $E(\text{host})$ is the total energy of the same supercell without the defect. The variables $n_i$ and $\mu_i$ are the number of species $i$ ($i$=In, P, and Bi) removed from the supercell in forming the defect and the corresponding chemical potential, respectively. $E_F$ is the Fermi level referenced to the valence band maximum (VBM, $E_{VBM}$) of the host. The donor transition level referenced to the CBM was calculated as [29]:

$$\varepsilon(0/q) = E_{CBM} + [E_{tot}(X,q) - E_{tot}(X,0)]/q.$$

**III. RESULTS AND DISCUSSION**

Before other magnetotransport measurements, the two point current-voltage (*I-V*) measurements were performed for each sample. Figure 1(a) shows the linear *I-V* curves for samples with $x$=1.47% and 2.15% at typical temperatures of 5 and 300 K. Similar linear *I-V* behavior was observed in all other samples, indicating the Ohmic contacts for all measured devices. Figure 1(b) shows the Hall resistance dependence of magnetic field for samples with $x$=1.47% and 2.15% when magnetic field was swept from -9 to 9 T, at the temperatures of 5, and 300 K. The Hall resistance linearly decreases with increasing the magnetic field not only for the two samples shown in Figure 1(b), but also for all other measured samples, indicating only one type of carrier need to be considered. The negative Hall coefficients prove the dominant *n*-type conductivity for all the investigated devices. Figure 1(c) shows the temperature dependence of the electron concentrations $n_e$ obtained from Hall Effect measurements for all the investigated InP$_{1-x}$Bi$_x$ samples with 0≤$x$≤2.41%. From 5 to 300 K, the electron concentrations of most InPBi samples show weak temperature dependence, only the two samples with high Bi compositions ($x$>2.15%) exhibit stronger temperature dependent $n_e$. In addition, for all the InPBi samples, large electron concentrations at low temperature of 5K are found in the range of $10^{17} \sim 10^{18}$ cm$^{-3}$. Especially, the undoped InP sample has the largest $n_e$ at 5 K. The high electron concentrations observed in all the investigated InPBi samples indicate degenerated gas formed in the conduction band, which is not originated from Bi. As an isovalent impurity in III-V semiconductors, Bi defect level is within the valence band [16,17], behaving as neither a good donor nor a good acceptor[30]. So, Bi should not consider as a charged impurity and has negligible contribution to the electron concentration, which leaves a problem as to the identification of the donors in the InPBi samples. The antisite P$_{In}$ defect is proposed to be the native donor in previous experiments for the *n*-type low temperature grown InP [19,20]. However, we find that the V$_P$ vacancy is another dominant donor, which will be discussed later.

The Bi composition dependence of the electron concentrations for all the InPBi samples at 5 and 300 K is shown in Figure 1(d). As we can see, the electron concentration non-monotonically varies with increasing the Bi doping. At low Bi compositions, the electron concentration decreases significantly from $3\times10^{18}$ to $2\times10^{17}$ cm$^{-3}$ with increasing $x$ from 0 to 1.83%. However, with increasing the Bi doping further the carrier concentration increases again as $x$ increasing. The Bi composition dependence of the electron concentration suggests that the incorporation of Bi-atoms will affect the properties of the native defects in InPBi, which in turn will result in different electron concentration.

Figure 2(a) shows the temperature dependent electric conductivity of a series of InP$_{1-x}$Bi$_x$ materials. Similar to the electron concentrations, the electric conductivity also exhibits weak temperature dependence except for the two highly doped samples. Using the obtained temperature dependence of conductivity and electron concentrations, the mobility of the InPBi samples can be obtain from $\mu(T)=\sigma(T)/(e\, n_e(T))$, where $\sigma$ is electric conductivity and $e$ is the elementary charge. The Bi concentration dependence of the electric mobility at temperatures of 5 and 300 K are show in Figure 2(b). For all the InPBi samples, the mobility at 300 K is higher than that of 5 K. For samples with low Bi compositions, mobility increases from around 600 to 1900 cm$^2$/Vs with increasing Bi doping up to 1.83%. With increasing Bi doping further, the electron mobility decreases with increasing Bi doping level. The Bi composition dependence of the mobility is in inverse with that of electron concentration, which is in accordance with the ionized impurity dominated scattering mechanism [31-35]. The decreasing of the electron concentration may due to the decreasing of the donor density or the increasing acceptor density, or vice versa. However, the effect of acceptors is negligible in our series of samples, with the main reasons discussed below. Firstly, the Hall resistance linearly decreases with increasing the magnetic field for all investigated samples, strongly indicating that the majority carriers in our samples are electrons, without noticeable compensation from acceptors. Secondly, increasing of the acceptor density will result in a decreasing of the electron concentration as well as the mobility, which is not in agreement with our experiment data. Thirdly, the most likely candidate acceptors in InPBi is suggested to be the Bi cluster defects [36], but this picture will result in a monotonic decreasing of the electron concentration as the increase of Bi compositions, disagree with our Hall measurements. Besides, similar to GaAsBi, the Bi related defects are expected to mainly affect the hole mobility rather than the electron mobility [37,38], and the hole concentration induced by these Bi defects is too low (in the order of $10^{14}$ cm$^{-3}$ when 0<$x$<4%) [39] to effectively compensate the high electron concentrations. Therefore we will only consider donors in our later discussions.

In a word, although the $n$-type conductivity originates from native donors rather than Bi itself, the nonlinear variations of electron concentration and mobility clearly suggest that the physical properties of the native donors are influenced by the Bi composition. In order to determine the dominant native donors, we perform first-principles calculations to study the native defects in bulk InP and InPBi alloys. We focus on two important quantities of the donors, the formation energy and the donor level, the former determining the concentration of the donors and the latter determining their ability to donate electrons to the host.

Figure 3 shows the formation energies of Bi and native defects at neutral state in bulk InP from our HSE calculations. The LT growth condition is supposed to be P rich or In poor (near $\mu_{In}$=-0.81eV in Figure 3). At this condition, Bi substituting P site has relatively low formation energy, in agreement with experiment that most Bi atoms are at substitutional sites [14]. The phosphorus antisite P$_{In}$ has the lowest formation energy among the native defects and the vacancy V$_P$ is the second most stable one. Both P$_{In}$ and V$_P$ can behave as donors and might be the candidates of the dominant donors in LT InP and InPBi. On the other hand, the In related native defects, In$_P$ and V$_{In}$, have higher formation energies and lower defect concentrations. Besides, their donor levels are quite deep within the gap [40]. Therefore, these two defects should have little effect on the electron concentration.

In previous experimental studies, the $n$-type conductivity of LT InP is explained by the prevailing P$_{In}$ antisite defects [19,20]. A donor level at 0.12 eV above CBM is determined to be the first donor transition level of P$_{In}$, while a deep donor level at 0.23 eV below CBM is assigned to be the second donor transition level [19,20]. Therefore, the $P_{In}$ antisite defects are expected to contribute high electron concentration in LT InP due to auto-ionization [19,20]. Considering the similarity of the growth conditions, we first assume the $P_{In}$ antisite as the dominant donor defect in our InPBi samples. However, we encounter two difficulties with this assumption: the $P_{In}$ does not have a donor level above CBM and cannot explain the nonlinear variation of electron concentration in InPBi as Bi composition increases. Based on our calculations, we find that $V_P$ is also a dominant donor and can solve the above difficulties, as will be discussed in details.

We first discuss the two difficulties with $P_{In}$ defect. From HSE calculations, the first and the second transition levels of the $P_{In}$ antisite are too deep within the gap, i.e., $\varepsilon(0/1+)=0.263$ and $\varepsilon(1+/2+)=0.295$ eV below CBM (Table I). Although the second transition level is in good agreement with the experiment, the first transition level has an opposite sign to the experimental value [19,20]. And we find that our calculations are in consistent with previous theoretical studies, where a deep first transition level of $P_{In}$ is obtained from various theoretical calculations [22,23]. In other words, $P_{In}$ is predicted to be a deep donor rather than an auto-ionized shallow donor, similar to the EL2 center of As antisite in GaAs [41].

The second difficult is that the $P_{In}$ antisite alone cannot explain the nonlinear variation of the electron concentration as the increase of Bi composition in InPBi. We calculate the formation energy and transition level of $P_{In}$ in InPBi alloys at different Bi compositions ($x$=0.93, 1.85, 2.77 and 3.70%). Different configurations of $P_{In}$ and Bi complexes are considered in our calculations. We use Bi$_n$-$P_{In}$ to denote the local atomic arrangement of $P_{In}$, where $n$ is the number of Bi nearest neighbors bonding to $P_{In}$. For higher compositions at $x$=2.77 and 3.70%, we only consider the Bi$_3$-$P_{In}$ and Bi$_4$-$P_{In}$ configurations, respectively. As shown in Figure 4(a), the formation energy of $P_{In}$ in the alloy highly depends on the number of Bi nearest neighbors $n$, but insensitive to the Bi composition $x$. As $n$ increases, the formation energy of $P_{In}$ decreases significantly by about 0.5 eV per Bi neighbor. This can be easily understood by strain effect, since the small size of $P_{In}$ can effectively release part of the local strain around the large Bi atoms. As the increase of Bi composition, $P_{In}$ antisite should have a higher chance to form $Bi_n$-$P_{In}$ complex, and the lower formation energy would result in an increasing $P_{In}$ defect concentration as the increasing $x$. Besides, the transition level of $P_{In}$ in the alloy is found to be slightly reduced with the increasing $x$ (Table I), mainly due to the band gap reduction effect and the associated decreasing of CBM induced by the incorporation of Bi. Therefore, as the increasing of Bi composition, the $P_{In}$ antisite is expected to have a higher defect concentration and a slightly lower transition level, leading to an increasing electron concentration as the increase of $x$, disagree with our Hall measurements.

Considering the two difficulties, we speculate there might be other dominant donor defects, that have a transition level above CBM and competes with $P_{In}$ in electron donation, leading to a nonlinear variation in electron concentration. To determine the possible donor defect, we calculate the donor transition levels for all the native defects. We find only P vacancy $V_P$ has a single particle defect level above CBM, and the donor level of $V_P$ is above CBM at 0.07 and 0.004 eV from PBE and HSE calculations, respectively, very close to the experimental value of 0.12 eV [19,20]. Our calculation on $V_P$ is also in good agreement with previous theoretical studies that predicted a shallow donor level of $V_P$ lying within 0.1 eV above CBM [40,42]. Considering that $V_P$ is the second most stable native defect (Figure 3), $V_P$ could also have a high defect concentration in both LT InP and InPBi. And the $V_P$ defects have indeed been observed in the scan tunneling microscope images in InPBi samples [14].

We further calculate the formation energy and donor level of $V_P$ in InPBi alloys. Similar to $P_{In}$, different $Bi_m$-$V_P$ complexes are considered, where $m$ denotes the number of Bi atoms that belong to the same tetrahedron as $V_P$. As seen in Table I, the donor levels of $V_P$ in the alloy from HSE calculations are above or very close to CBM, and the values from PBE are all above CBM (not shown). Considering the error of transition level is in the order of 0.1 eV from first-principles calculations, we believe that $V_P$ in the alloy can maintain its shallow donor nature as in InP. From Figure 4(b), the formation energy of $V_P$ in InPBi also depends on its local configuration and is decreased with increasing $m$ in a non-monotonic manner. The decreasing in the formation energy is due to the partial release of the strain around Bi atoms, and the non-monotonic variation is due to the complicated structural distortion around the Bi$_m$-V$_P$ complexes, where Bi$_m$ cluster tends to push the center In atom towards to the vacancy. The distortion not only depends on the number $m$, but also on the second neighbor or even further Bi atoms. For example, for a Bi$_4$ tetrahedral cluster, the V$_P$ on the neighboring P site could form a local Bi$_1$-V$_P$ or Bi$_2$-V$_P$ complexes, and the former has a lower energy than the latter by 0.10 eV. Nevertheless, from Figure 4(b) we expect the formation energy of V$_P$ is decreased if V$_P$ forms complexes with Bi at high Bi composition. However, since Bi also occupies P site and has much lower formation energy (Figure 3), a competition between Bi and V$_P$ is expected. We study a Bi atom occupying the V$_P$ site in bulk InP and obtain a negative formation energy of -2.05 eV, indicating this is a exothermic process. Therefore, Bi will occupy part of the V$_P$ sites in the alloys. At low and intermediate Bi compositions, this can effectively reduce the V$_P$ defect concentration. At higher Bi composition, Bi can form clusters and the formation energy of V$_P$ is reduced due to the formation of Bi$_m$-V$_P$ complexes. As a result, the V$_P$ defect concentration might increase again with increasing $x$ at high Bi compositions.

With the above analysis, we reach the following picture of the nonlinear variation of the electron concentration with increasing the Bi content in InPBi alloys. We propose that both of P$_{In}$ and V$_P$ are the dominant donors in InP and InPBi alloys. The donor level 0.23 eV below CBM observed in experiment is the donor level of P$_{In}$, while the one 0.12 eV above CBM is essentially the first donor level of V$_P$. At low Bi composition ($x$<1.83%), the reduction of V$_P$ defect concentration is the dominant effect that governs the variation of electron concentration, i.e., electron concentration decreases with increasing $x$. At higher Bi composition ($x$>1.83%), the V$_P$ concentration is increased, and the electrons contributed by P$_{In}$ donors would be increased significantly due to its lower formation energy and smaller donor level. Thus the electron concentration will be increased abruptly as the increasing $x$. This picture can also explain the inverse variation of the electron mobility. When the impurity scattering mechanism is the dominant mechanism for the conduction band transport, the mobility first increases mainly due to the reduction of $V_P$ defect density when $x<1.83\%$, and then decreases again due to the increase of $P_{In}$ and $V_P$ defect densities.

We now can have a better understanding of the temperature dependence of the electron concentration (Figure 1(c)). Since $V_P$ has a donor level above CBM, $V_P$ is fully ionized even at 5 K. The electrons at this low temperature are mainly contributed by $V_P$, and the variation of $n_e$ as a function of Bi composition also manifests the variation of $V_P$ defect density. As the temperature increases, $P_{In}$ contributes more electrons to the InPBi host. However, $P_{In}$ is far from fully activation due to its deep donor nature. Thus most InPBi samples show weak temperature dependence up to 300K. Only at high Bi compositions ($x>2.18\%$), $P_{In}$ has a higher defect density by forming $Bi_n$-$P_{In}$ complexes, and has a higher contribution to the electron concentration. As a result, we expect stronger temperature dependence of $n_e$ and a larger difference of $n_e$ between 5 K and 300 K, as shown in Figures 1(c) and 1(d). Besides, we also observe a fluctuation of $n_e$ at 300 K at high Bi composition [Figure 1(d)]. This indicates a fluctuation of $P_{In}$ defect density, probably due to compositional inhomogeneity at high Bi composition.

## IV. CONCLUSIONS

Through varied temperature magnetotransport measurements in Hall-bar devices, we find that the InPBi alloys have very high electron concentrations that vary nonlinearly with the Bi composition. Combined with first-principles calculations, we determine that both $P_{In}$ and $V_P$ are the dominant donors in the alloys. At low Bi composition $x<1.83\%$, the reduction of $V_P$ defect concentration is the main effect to govern the decreasing of electron concentration and the increasing of mobility. At higher Bi composition, the increasing defect concentrations of $P_{In}$ and $V_P$ leads to a significant increasing in the electron concentration and the reversed variation in mobility. We present the electrical transport properties of InPBi alloys, and reveal the mechanism by changing the Bi composition to tune the electron concentration and mobility in InPBi and other dilute Bi alloys.

ACKNOWLEDGMENTS

We thank Prof. S.-H. Wei and Prof. J. W. Luo for useful discussion and suggestions. This work was supported by '973 Program' No. 2014CB643900, and NSFC Grant Nos. 61225021, 11474272, 11204296, and 11474247.

**FIGURES:**

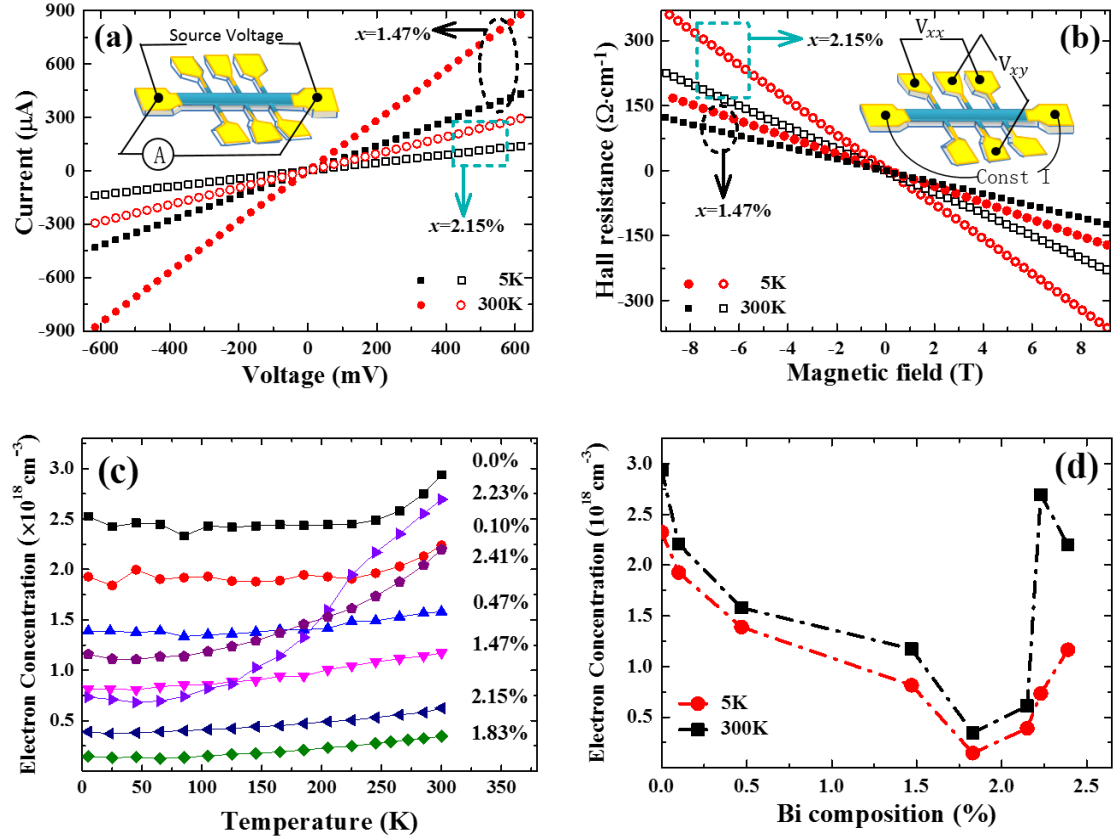

Figure 1. (a) The *I*(*V*) curves and (b) the magnetic field dependent Hall resistances R$_{xy}$ for InP$_{1-x}$Bi$_x$ samples with *x*=1.47% (solid symbols) and 2.15% (open symbols), measured at the typical temperatures of 5 and 300 K. (c) The temperature dependence of the electron concentrations for InP$_{1-x}$Bi$_x$ samples with 0≤*x*≤2.41%. (d) Bi composition dependent electron concentration of a series of InP$_{1-x}$Bi$_x$ alloys at 5 and 300 K, respectively.

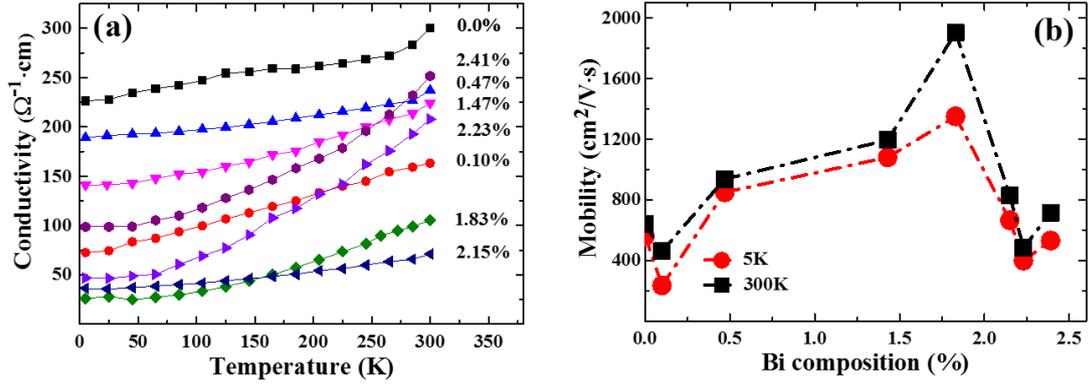

Figure 2. (a) Temperature dependent electric conductivity of a series of InP$_{1-x}$Bi$_x$ epilayers with 0≤$x$≤2.41%. (b) Bi composition dependent electron mobility of a series of InP$_{1-x}$Bi$_x$ alloys at 5 and 300K, respectively.

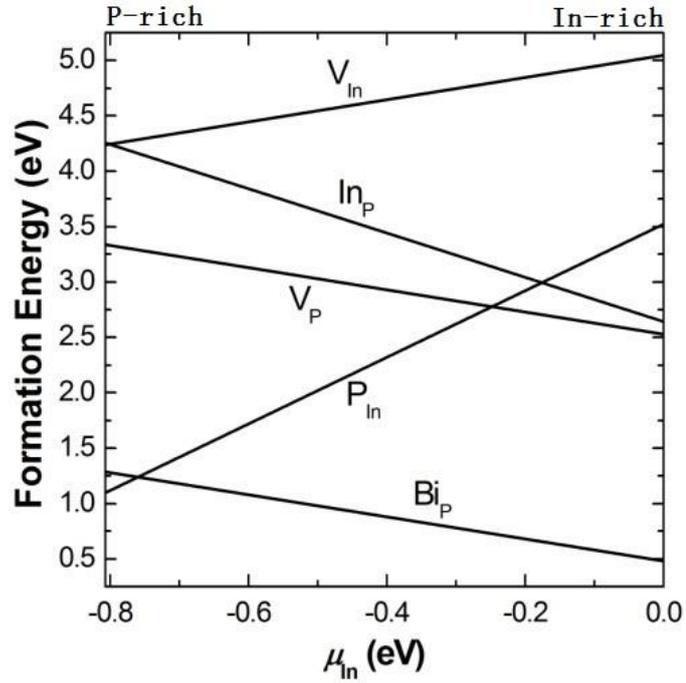

Figure 3. Formation energy of neutral native defects and BiP defect in bulk InP at various In chemical potentials from HSE.

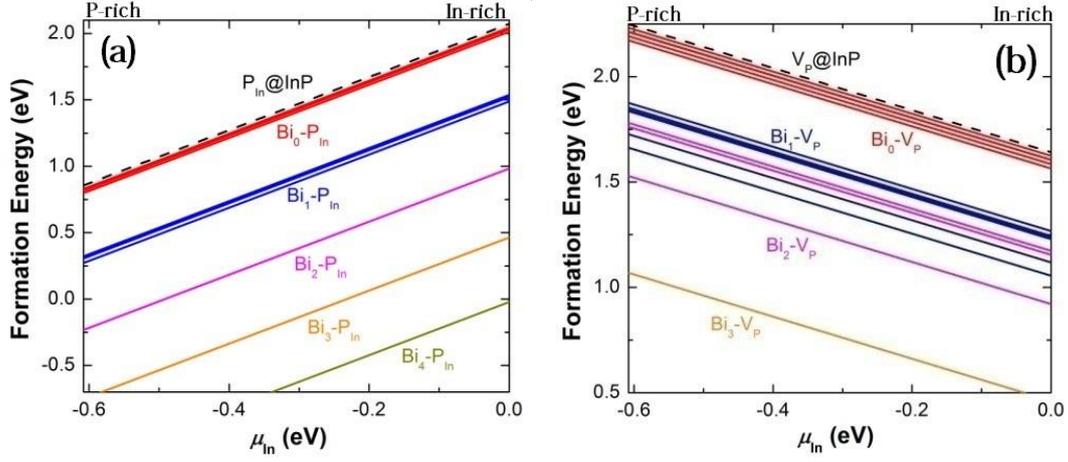

Figure 4. Formation energies of (a) $P_{In}$ and (b) $V_P$ in InPBi alloys at different Bi compositions from HSE. $Bi_n$-$P_{In}$ and $Bi_m$-$V_P$ denote the local structure of defect with respective to Bi clusters. The defect formation energies in bulk InP are also shown with dashed lines for comparison.

**TABLE:**

TABLE I. Transition donor levels with respect to CBM of $P_{In}$ and $V_P$ in bulk InP and InPBi alloys (unit in eV) from HSE. The positive and negative values correspond to donor levels below and above CBM, respectively.

|  | $x$=0% | $x$=0.83% | $x$=0.83% | $x$=1.85% | $x$=2.77% | $x$=3.70% |
|---|---|---|---|---|---|---|
| **$P_{In}$** | InP bulk | $Bi_0$-$P_{In}$ | $Bi_1$-$P_{In}$ | $Bi_2$-$P_{In}$ | $Bi_3$-$P_{In}$ | $Bi_4$-$P_{In}$ |
| $\varepsilon(0/1+)$ | 0.263 | 0.235 | 0.214 | 0.206 | 0.186 | 0.167 |
| $\varepsilon(1+/2+)$ | 0.295 | 0.260 | 0.260 | 0.227 | 0.200 | 0.175 |
| **$V_P$** | InP bulk | $Bi_0$-$V_P$ | $Bi_1$-$V_P$ | $Bi_2$-$V_P$ | $Bi_3$-$V_P$ | $Bi_4$-$V_P$ |
| $\varepsilon(0/1+)$ | -0.004 | 0.008 | -0.018 | -0.005 | 0.034 | - |